\documentclass[prd,twocolumn,floatfix,amsmath,nofootinbib,amssymb,floatfix]{revtex4}
\usepackage{graphicx,color,dcolumn,booktabs,bm}
\usepackage{longtable,lscape}
\usepackage{txfonts}
\usepackage{overpic}
\usepackage{amssymb}
\usepackage{indentfirst}
\usepackage{feynmf}   
\usepackage{slashed}  
\usepackage{cases}
\usepackage{color}
\usepackage{multirow}
\usepackage{epstopdf}
\usepackage{enumerate}
\usepackage{graphicx,color,dcolumn,booktabs,bm}
\usepackage[colorlinks, citecolor=blue,anchorcolor=red,menucolor=red, linkcolor=red,filecolor=red,urlcolor=blue,frenchlinks=red]{hyperref}

\graphicspath{{Figures/}} %

\begin{document}

\title{Deciphering the light vector meson contribution to the cross sections of $e^+e^-$ annihilations into the open-strange channels through a combined analysis} 
\author{Jun-Zhang Wang$^{1,2}$}\email{wangjzh2012@lzu.edu.cn}
\author{Li-Ming Wang$^{3}$}\email{lmwang@ysu.edu.cn}
\author{Xiang Liu$^{1,2,4}$\footnote{Corresponding author}}\email{xiangliu@lzu.edu.cn}
\author{Takayuki Matsuki$^5$}\email{matsuki@tokyo-kasei.ac.jp}
\affiliation{$^1$School of Physical Science and Technology, Lanzhou University, Lanzhou 730000, China\\
$^2$Research Center for Hadron and CSR Physics, Lanzhou University $\&$ Institute of Modern Physics of CAS, Lanzhou 730000, China\\
$^3$Key Laboratory for Microstructural Material Physics of Hebei Province, School of Science, Yanshan University, Qinhuangdao 066004, China\\
$^4$Lanzhou Center for Theoretical Physics, Key Laboratory of Theoretical Physics of Gansu Province, and Frontiers Science Center for Rare Isotopes, Lanzhou University, Lanzhou 730000, China\\
$^5$Tokyo Kasei University, 1-18-1 Kaga, Itabashi, Tokyo 173-8602, Japan}

\date{\today}

\begin{abstract}
In this work, we perform a combined 
analysis to the measured data
of the cross section of open-strange processes $e^+e^- \to K^+K^-$, $e^+e^- \to K\bar{K}^{*}+c.c.$, $e^+e^- \to K^{*+}K^{*-}$, $e^+e^- \to K_1(1270)^+K^-$, $e^+e^- \to K_1(1400)^+K^-$, $e^+e^- \to K_2^{*}(1430)\bar{K}+c.c.$ and $e^+e^- \to K(1460)^+K^-$ with the support of study of hadron spectroscopy. We reveal the contribution of the possible light vector mesons around 2 GeV to reproduce the cross section data of the reported open-strange processes from $e^+e^-$ annihilation which may provide a new perspective to construct the light vector meson family and understand the $Y(2175)$.

\end{abstract}

\maketitle

\section{Introduction}

Recently, the BESIII Collaboration released the data of the cross section for the open-strange process $e^+e^- \to K^+K^-$, where an obvious enhancement at 2.2 GeV was observed and its Breit-Wigner resonance parameters were measured to be $M=2239.2 \pm 7.1 \pm 11.3$  MeV and $\Gamma=139.8 \pm 12.3 \pm 20.6$  MeV \cite{Ablikim:2018iyx}. This is the first evidence showing the observation of the vector structure around 2.2 GeV in the open-strange channel. After that, BESIII reported the partial-wave analysis result of a new open-strange process $e^+e^- \to K^+K^-\pi^0 \pi^0$, where the Born cross sections for the subprocesses $e^+e^- \to K^+(1460)K^-, K^+_1(1400)K^-, K^+_1(1270)K^- $, and $ K^{*+}K^{*-}$ were measured \cite{Ablikim:2020pgw}. By performing an overall fit to the above four processes, a structure with the mass of $2126.5 \pm 16.8 \pm 12.4$ MeV and the width of $106.9 \pm 32.1 \pm 28.1$ MeV was found, although a limited significance appears in the process $e^+e^- \to K^+_1(1270)K^-$  and $ K^{*+}K^{*-}$ \cite{Ablikim:2020pgw}. These new open-strange processes again indicate the existence of a vector structure around 2.2 GeV. When checking the data collected by the Particle Data Group (PDG) \cite{Zyla:2020zbs}, one finds that the resonance parameter of this structure observed in these open-strange processes is close to the corresponding average value of that of the $Y(2175)$.  Therefore, such simple comparison may suggest that this newly observed vector structure around 2.2 GeV and the $Y(2175)$ are the same state.

As we all know, the $Y(2175)$ was first reported in the process $e^+e^-\to \phi f_0(980)$ by the BaBar Collaboration based on the initial-state-radiation (ISR) method \cite{Aubert:2006bu}, which was later confirmed in the same process by Belle \cite{Shen:2009zze} and in the process $J/\psi \to\eta\phi f_0(980)$ by BES \cite{Ablikim:2007ab} and BESIII \cite{Ablikim:2014pfc}.
Different from most of the observed charmoniumlike $Y$ states, the $Y(2175)$ is the only light-flavor $Y$ particle, which was reported in the hidden-strange final states and still remains a mystery even though more than a decade have passed since its discovery. Thus, it has inspired theorist's great interest in exploring its inner structure. In the past years, the theoretical explanations for the $Y(2175)$ include hybrid $s\bar{s}g$ \cite{Ding:2006ya}, vector strangeonium state $\phi(3S)$ \cite{Barnes:2002mu,Pang:2019ttv} and $\phi(2D)$ \cite{Wang:2012wa,Ding:2007pc,Pang:2019ttv}, which should favor the open-strange decay modes. 
However, the recent BESIII measurement of the $e^+e^- \to K^+K^-$ \cite{Ablikim:2018iyx} and $e^+e^- \to K^+K^-\pi^0 \pi^0$ \cite{Ablikim:2020pgw} reactions indicates that it is hard to understand these open-strange experimental data under the hybrid and strangeonium assignments to the $Y(2175)$. 

If the $Y(2175)$ is the $s\bar{s}g$ hybrid state, the dominant decay modes should be $K_1(1270)^+K^-$ and $K_1(1400)^+K^-$ \cite{Ding:2006ya}, while two open-strange decay channels like $K^+K^-$ and $K^{*+}K^{*-}$ 
should be strongly suppressed 
according to the flux tube model analysis \cite{Close:1994hc,Close:2003mb} and the QCD sum rule calculation \cite{Zhu:1998sv,Zhu:1999wg}. On the other hand, different predictions for the branching ratios of the open-strange decay modes of strangeonium states $\phi(3S)$ and $\phi(2D)$ were given in former theoretical studies. For example, the branching ratio of $\phi(3S) \to K^+K^-$ was calculated to be almost zero in Ref. \cite{Barnes:2002mu}, and instead, $K_1(1270)^+K^-$  and $ K^{*+}{K}^{*-}$ were predicted to be dominant decay channels of $\phi(3S)$ and $\phi(2D)$ in Refs. \cite{Barnes:2002mu,Ding:2007pc,Pang:2019ttv}.  Furthermore, the theoretical total widths of $\phi(3S)$ and $\phi(2D)$ are generally estimated to be $200 \sim 400$ MeV \cite{Barnes:2002mu,Pang:2019ttv,Wang:2012wa,Ding:2007pc}, which obviously deviates from the present experimental data of the $Y(2175)$ \cite{Zyla:2020zbs}. 
Therefore, the BESIII data \cite{Ablikim:2018iyx,Ablikim:2020pgw} does not support the above theoretical predictions, which becomes a big challenge for the $s\bar{s}$ or $s\bar{s}g$ assignment to the $Y(2175)$. 
In addition to the above difficulties in decoding the $Y(2175)$ structure, there are still two confusing problems, which should be mentioned here. Through a comparison among the cross sections of the reported open-strange processes from $e^+e^-$ annihilation,  we can see that although the resonance parameter of the observed vector structure around 2.2 GeV in $e^+e^- \to K^+K^-$ \cite{Ablikim:2018iyx} is similar to that in $e^+e^- \to K^+K^-\pi^0 \pi^0$ \cite{Ablikim:2020pgw}, there is still about 100 MeV deviation on these measured masses. Hence, this measured mass discrepancy problem should be clarified. In addition, the BESIII measurement shows no observation of the structure around 2.2 GeV in a typical open-strange $K^{*}K^{*}$ mode \cite{Ablikim:2020pgw}, which also should be appropriately understood.

When facing this puzzling situation, let us firstly return to the open-strange reaction itself $e^+e^- \to K^{(*/\prime)}_{(J)}\bar{K}^{(*/\prime)}_{(J)}$ ($e^+e^- \to K^+K^-$, $e^+e^- \to K\bar{K}^{*}+c.c.$, $e^+e^- \to K^{*+}K^{*-}$, $e^+e^- \to K_1(1270)^+K^-$, $e^+e^- \to K_1(1400)^+K^-$, $e^+e^- \to K_2^{*}(1430)\bar{K}+c.c.$ and $e^+e^- \to K(1460)^+K^-$).  
In fact, the open-strange process from $e^+e^-$ annihilation at center-of-mass energy $\sqrt{s}\sim 2$ GeV is mediated by different intermediate vector mesons. Since there is no isospin restriction, the highly excited $\rho$, $\omega$, and $\phi$ meson states around 2 GeV may have contributions to the cross sections of the discussed open-strange processes, which makes the whole analysis 
quite complicated. Therefore, when facing the enhancement or dip structure around 2.2 GeV observed in both $e^+e^- \to K^+K^-$ and $e^+e^- \to K^+K^-\pi^0 \pi^0$ \cite{Ablikim:2018iyx,Ablikim:2020pgw}, we see that it is a rough treatment to consider only a simple Breit-Wigner fit to 
the cross section data 
of the discussed open-strange processes from $e^+e^-$ annihilation \cite{Ablikim:2018iyx,Ablikim:2020pgw,BaBar:2013jqz,Aubert:2007ym}, which may inevitably lead to a
puzzling situation mentioned above.


Obviously, we need to carry out a combined analysis to the present measured data of cross sections producing open-strange channels via $e^+e^-$ annihilation \cite{Ablikim:2018iyx,Ablikim:2020pgw,BaBar:2013jqz,Aubert:2007ym}, which is supported by the study of hadron spectroscopy. In fact, this approach was applied to construct higher charmonia above 4 GeV \cite{Wang:2019mhs,Wang:2020prx,Wang:2018rjg}. In this work, the hadron spectrum analysis is based on an unquenched potential model \cite{Pang:2017dlw,Song:2015nia,Song:2015fha,Wang:2016krl,Wang:2019mhs,Wang:2020prx,Wang:2018rjg}, by which 
we select suitable light vector mesons with mass around 2 GeV involved in the discussed processes. Of course, this mass spectrum analysis simultaneously provides the numerical spatial wave functions of the selected light vector mesons, which are applied to calculate their two-body open-strange decay widths and di-lepton widths. With this preparation supported by the meson spectroscopy, the main task of the present work is to perform a combined analysis to the experimental cross section data of $e^+e^-$ annihilations into open-strange channels, which can 
provide valuable information of light vector meson contribution to depict 
the experimental cross section data of $e^+e^-$ annihilation into open-strange channels. Finally, we show later that the above puzzling situation can be clarified, which also provides a new perspective to  construct the light vector meson family and understand the $Y(2175)$.

This paper is organized as follows. After Introduction, we discuss the cross section of open-strange processes $e^+e^- \to K^{(*/\prime)}_{(J)}\bar{K}^{(*/\prime)}_{(J)}$ by considering the direct production and light vector meson contributions in Sec. \ref{sec2}. Furthermore, we illustrate how to obtain the mass spectrum and decay behaviors of light vector mesons in Sec. \ref{sec3}. In Sec. \ref{sec4}, based on our theoretical predictions for the branching ratios of the open-strange strong decay modes and di-lepton widths of  higher light vector mesons above 2 GeV, we find that higher light vector meson states play an important role to perform a combined analysis of the cross sections of $e^+e^-$ annihilations into the open-strange processes. Finally, this work ends with a summary in Sec. \ref{sec5}.

\section{ cross sections of open-strange processes $e^+e^- \to K^{(*/\prime)}_{(J)}\bar{K}^{(*/\prime)}_{(J)}$ around 2.0 GeV}\label{sec2}

\begin{figure}[b]
	\includegraphics[width=8.5cm,keepaspectratio]{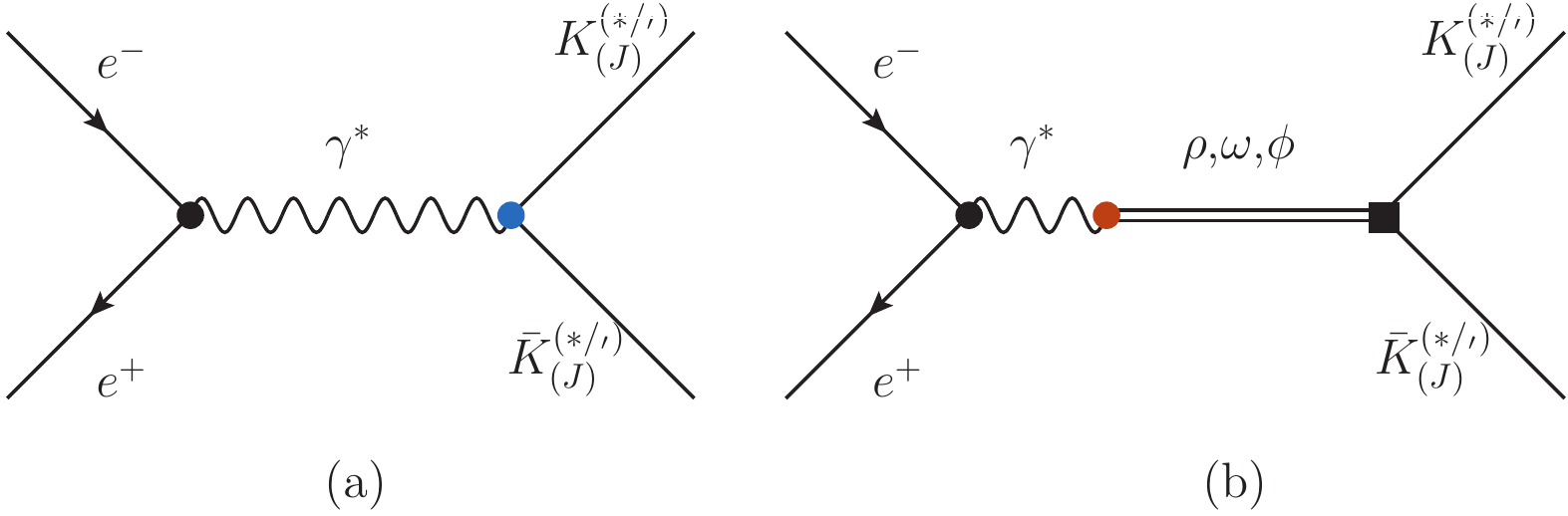}
	\caption{ The schematic diagrams for depicting the open-strange process $e^+e^- \to K^{(*/\prime)}_{(J)}\bar{K}^{(*/\prime)}_{(J)}$. Diagram (a) corresponds to a direct annihilation process, while diagram (b) is the resonance contributions from the intermediate $\rho$, $\omega$, $\phi$ state. }\label{Feyndiagram}
\end{figure}

In this section, we focus on the open-strange reactions $e^+e^- \to K^+{K}^-$ \cite{Ablikim:2018iyx,BaBar:2013jqz}, $e^+e^- \to K\bar{K}^{*}+c.c.$ \cite{Aubert:2007ym}, $e^+e^- \to K^{*+}K^{*-}$ \cite{Ablikim:2020pgw}, $e^+e^- \to K_1(1270)^+K^-$ \cite{Ablikim:2020pgw}, $e^+e^- \to K_1(1400)^+K^-$ \cite{Ablikim:2020pgw}, $e^+e^- \to K_2^{*}(1430)\bar{K}+c.c.$ \cite{Aubert:2007ym} and $e^+e^- \to K(1460)^+K^-$ \cite{Ablikim:2020pgw} with center-of-mass (CM) energy around 2.0 GeV and illustrate how to prensent the cross section. In addition to the $K^*K^*$, $K_1(1270)K$, $K_1(1400)K$ and $ K(1460)K$ channels measured by BESIII  for the first time,  there also exist the experimental results of  several other open-strange channels, such as  $KK^*$ and $K_2(1430)^*K$, where their measured results were reported by the BaBar Collaboration in 2008 \cite{Aubert:2007ym}.   As shown in Fig. \ref{Feyndiagram}, there exist two mechanisms working together for the $e^+e^- \to K^{(*/\prime)}_{(J)}\bar{K}^{(*/\prime)}_{(J)}$ process.  The first one is the direct annihilation of $e^+e^-$ into open-strange channel, where the virtual photon directly couples with the final states $K^{(*/\prime)}_{(J)}\bar{K}^{(*/\prime)}_{(J)}$, which provides the background contribution. The second one occurs via the intermediate light vector meson states, which include the $\rho,\omega$, and $\phi$-like resonances.

In this work, the effective Lagrangian approach is adopted to calculate the discussed open-strange process $e^+e^- \to K^{(*/\prime)}_{(J)}\bar{K}^{(*/\prime)}_{(J)}$ as shown in Fig. \ref{Feyndiagram}. The effective Lagrangian densities involved in the concrete works include \cite{Bauer:1975bv,Bauer:1975bw,Chen:2011cj,Huang:2013jda,Xie:2015wja}
\begin{eqnarray}
\mathcal{L}_{KK\gamma}&=&ieA^{\mu}(\bar{K}\partial_{\mu}K-\partial_{\mu}\bar{K} K),\nonumber\\
\mathcal{L}_{KK^*\gamma}&=&e\varepsilon^{\mu\nu\rho\sigma}\partial_{\mu}A_{\nu}(\bar{K}\partial_{\rho}K^*_{\sigma}+\partial_{\rho}\bar{K}^*_{\sigma} K),\nonumber\\
\mathcal{L}_{K^*K^*\gamma}&=&ie(A^{\mu}(\bar{K}^*_{\nu}\overleftrightarrow{\partial}_{\mu}K^{*\nu})+\bar{K}^{*\mu}(K^*_{\nu}\overleftrightarrow{\partial}_{\mu}A^{\nu}) \nonumber\\
&&+(A^{\nu}\overleftrightarrow{\partial}_{\mu}\bar{K}^*_{\nu})K^{*\mu}), \nonumber\\
\mathcal{L}_{KK_1\gamma}&=&ie A_{\mu}(\bar{K}K_1^{\mu}-\bar{K}_1^{\mu} K),\nonumber\\
\mathcal{L}_{KK_2\gamma}&=&e\varepsilon_{\mu\nu\rho\sigma}\partial^{\rho}A^{\sigma}(\partial_{\delta}\bar{K}\partial^{\mu}K_2^{\nu\delta}
+\partial^{\mu}\bar{K}_2^{\nu\delta}\partial_{\delta}K),\nonumber\\
\mathcal{L}_{KK^{\prime}\gamma}&=&ieA^{\mu}(\bar{K}\overleftrightarrow{\partial}_{\mu}K^{\prime}+c.c.), \\
\mathcal{L}_{\gamma \mathcal{V} }&=&\frac{-em_{\mathcal{V} }^2}{f_{\mathcal{V} }}\mathcal{V} _{\mu}A^{\mu}, \nonumber \\
\mathcal{L}_{\mathcal{V}  KK}&=&i g_{\mathcal{V}  KK}(\bar{K}\partial_{\mu}K-\partial_{\mu}\bar{K} K)\mathcal{V} ^{\mu},\nonumber\\
\mathcal{L}_{\mathcal{V}  KK^*}&=&g_{\mathcal{V} KK^*}\varepsilon^{\mu\nu\rho\sigma}(\bar{K}\partial_{\rho}K^*_{\sigma}+\partial_{\rho}\bar{K}^*_{\sigma} K)\partial_{\mu}\mathcal{V} _{\nu},\nonumber
\end{eqnarray}
\begin{eqnarray}
\mathcal{L}_{\mathcal{V}  K^*K^*}&=&i g_{\mathcal{V}  K^*K^*}((\bar{K}^*_{\nu}\overleftrightarrow{\partial}_{\mu}K^{*\nu})\mathcal{V} ^{\mu}+\bar{K}^{*\mu}(K^*_{\nu}\overleftrightarrow{\partial}_{\mu}\mathcal{V} ^{\nu}) \nonumber\\
&&+(\mathcal{V} ^{\nu}\overleftrightarrow{\partial}_{\mu}\bar{K}^*_{\nu})K^{*\mu}), \nonumber\\
\mathcal{L}_{\mathcal{V}  KK_1}&=&i g_{\mathcal{V}  KK_1}(\bar{K}K_1^{\mu}-\bar{K}_1^{\mu} K)\mathcal{V} _{\mu},\nonumber\\
\mathcal{L}_{\mathcal{V}  KK_2}&=&g_{\mathcal{V}  KK_2}\varepsilon_{\mu\nu\rho\sigma}(\partial_{\delta}\bar{K}\partial^{\mu}K_2^{\nu\delta}
+\partial^{\mu}\bar{K}_2^{\nu\delta}\partial_{\delta}K)\partial^{\rho}\mathcal{V}^{\sigma}, \nonumber\\
\mathcal{L}_{\mathcal{V}  KK^{\prime}}&=&i g_{\mathcal{V}  KK^{\prime}}(\bar{K}\overleftrightarrow{\partial}_{\mu}K^{\prime}+c.c.)\mathcal{V} ^{\mu},
\end{eqnarray}
where $K_1$, $K_2$, and $K^{\prime}$ stand for the kaon meson fields of $K_1(1270)/K_1(1400)$, $K_2^*(1430)$ and $K(1460)$, respectively. And, $\mathcal{V}$ is the intermediate light vector meson field. $g_{\mathcal{V}  K^{(*/\prime)}_{(J)}K^{(*/\prime)}_{(J)}}$ is the corresponding coupling constants involved in the $\mathcal{V}$ state and the open-strange channel.

The scattering amplitudes of seven open-strange processes $e^+e^- \to KK$, $e^+e^- \to KK^{*}$, $e^+e^- \to K^{*}K^{*}$, $e^+e^- \to K_1(1270)K$, $e^+e^- \to K_1(1400)K$, $e^+e^- \to K_2^{*}(1430)K$, $e^+e^- \to K(1460)K$ generally depicted in in Fig. \ref{Feyndiagram} can be written as
\begin{eqnarray}
\mathcal{M}_{\mathrm{Dir}}^{KK^{(\prime)}}&=&[\bar{v}(k_2)ie\gamma^{\mu}u(k_1)]\frac{-g_{\mu\nu}}{q^2}[-e(p_4^{\nu}-p_3^{\nu})F_{KK^{(\prime)}}(q^2)], \nonumber \\
\mathcal{M}_{\mathrm{\mathcal{V} }}^{KK^{(\prime)}}&=&[\bar{v}(k_2)ie\gamma_{\mu}u(k_1)]\frac{-g^{\mu\rho}}{q^2}\frac{-em_{\mathcal{V} }^2}{f_{\mathcal{V} }}\frac{-g_{\rho\nu}+q_{\rho}q_{\nu}/m_{\mathcal{V} }^2}{q^2-m_{\mathcal{V} }^2+i m_{\mathcal{V} }\Gamma_{\mathcal{V} }}\nonumber\\
&&\times[- g_{\mathcal{V} KK^{(\prime)}}(p_4^{\nu}-p_3^{\nu})],  \label{eq22} \\
\mathcal{M}_{\mathrm{Dir}}^{KK^*}&=&[\bar{v}(k_2)ie\gamma^{\mu}u(k_1)]\frac{-g_{\mu\nu}}{q^2}[e\varepsilon^{\alpha\nu\rho\sigma}q_{\alpha}
p_{4\rho}\epsilon_{K^*\sigma}^{ *}F_{KK^*}(q^2)], \nonumber \\
\mathcal{M}_{\mathrm{\mathcal{V} }}^{KK^*}&=&[\bar{v}(k_2)ie\gamma_{\mu}u(k_1)]\frac{-g^{\mu\rho}}{q^2}\frac{-em_{\mathcal{V} }^2}{f_{\mathcal{V} }}\frac{-g_{\rho\nu}+q_{\rho}q_{\nu}/m_{\mathcal{V} }^2}{q^2-m_{\mathcal{V}}^2+i m_{\mathcal{V} }\Gamma_{\mathcal{V}}}\nonumber\\
&&\times[g_{\mathcal{V}  KK^*}\varepsilon^{\alpha\nu\omega\sigma}q_{\alpha}p_{4\omega}\epsilon_{K^*\sigma}^{ *}],  \\
\mathcal{M}_{\mathrm{Dir}}^{K^*K^*}&=&[\bar{v}(k_2)ie\gamma^{\mu}u(k_1)]\frac{-g_{\mu\nu}}{q^2}[-e \Big(g^{\alpha\beta}(p_{4}^{\nu}-p_{3}^{\nu})-g^{\nu\beta}q^{\alpha} \nonumber \\
&&+g^{\nu\alpha}q^{\beta}+g^{\nu\alpha}p_{3}^{\beta}-g^{\nu\beta}p_{4}^{\alpha}\Big)\epsilon_{K^*\alpha}^{ *}\epsilon_{K^*\beta}^{ *} F_{K^*K^*}(q^2)], \nonumber \\
\mathcal{M}_{\mathrm{\mathcal{V}}}^{K^*K^*}&=&[\bar{v}(k_2)ie\gamma_{\mu}u(k_1)]\frac{-g^{\mu\rho}}{q^2}\frac{-em_{\mathcal{V}}^2}{f_{\mathcal{V}}}\frac{-g_{\rho\nu}+q_{\rho}q_{\nu}/m_{\mathcal{V}}^2}{q^2-m_{\mathcal{V}}^2+i m_{\mathcal{V}}\Gamma_{\mathcal{V}}}\nonumber\\
&&\times[-g_{\mathcal{V} K^*K^*}\Big(g^{\alpha\beta}(p_{4}^{\nu}-p_{3}^{\nu})-g^{\nu\beta}q^{\alpha}+g^{\nu\alpha}q^{\beta} \nonumber \\
&&+g^{\nu\alpha}p_{3}^{\beta}-g^{\nu\beta}p_{4}^{\alpha}\Big)\epsilon_{K^*\alpha}^{ *}\epsilon_{K^*\beta}^{ *}], \\
\mathcal{M}_{\mathrm{Dir}}^{KK_1}&=&[\bar{v}(k_2)ie\gamma^{\mu}u(k_1)]\frac{-g_{\mu\nu}}{q^2}[ie\epsilon_{K_1}^{\nu *}F_{KK_1}(q^2)], \nonumber \\
\mathcal{M}_{\mathrm{\mathcal{V}}}^{KK_1}&=&[\bar{v}(k_2)ie\gamma_{\mu}u(k_1)]\frac{-g^{\mu\rho}}{q^2}\frac{-em_{\mathcal{V}}^2}{f_{\mathcal{V}}}\frac{-g_{\rho\nu}+q_{\rho}q_{\nu}/m_{\mathcal{V}}^2}{q^2-m_{\mathcal{V}}^2+i m_{\mathcal{V}}\Gamma_{\mathcal{V}}}\nonumber\\
&&\times[i g_{\mathcal{V} KK_1}\epsilon_{K_1}^{\nu *}],  \\
\mathcal{M}_{\mathrm{Dir}}^{KK_2}&=&[\bar{v}(k_2)ie\gamma^{\mu}u(k_1)]\frac{-g_{\mu\nu}}{q^2}[ie\varepsilon^{\alpha\beta\rho\nu}p_3^{\delta}q_{\rho}
p_{4\alpha}\epsilon_{K_2\beta\delta}^{ *} \nonumber \\
&&\times F_{KK_2}(q^2)], \nonumber \\
\mathcal{M}_{\mathrm{\mathcal{V}}}^{KK_2}&=&[\bar{v}(k_2)ie\gamma_{\mu}u(k_1)]\frac{-g^{\mu\rho}}{q^2}\frac{-em_{\mathcal{V}}^2}{f_{\mathcal{V}}}\frac{-g_{\rho\nu}+q_{\rho}q_{\nu}/m_{\mathcal{V}}^2}{q^2-m_{\mathcal{V}}^2+i m_{\mathcal{V}}\Gamma_{\mathcal{V}}}\nonumber\\
&&\times[i g_{\mathcal{V} KK_2}\varepsilon^{\alpha\beta\omega\nu}p_3^{\delta}q_{\omega}
p_{4\alpha}\epsilon_{K_2\beta\delta}^{ *}], \label{eq26}
\end{eqnarray}
where $q^{\mu}=(\sqrt{s},0,0,0)$ and $p_3$ and $p_4$ are the four-momenta of final states $K^{(*/\prime)}_{(J)}$/$\bar{K}^{(*/\prime)}_{(J)}$. Additionally, the time-like form factor $F_{K^{(*/\prime)}_{(J)}K^{(*/\prime)}_{(J)}}(q^2)$ is introduced when depicting the direct production process. Generally speaking, the expression of the form factor in the time-like region is relatively complicated. For simplicity, we adopt a universal form factor $F_{K^{(*/\prime)}_{(J)}K^{(*/\prime)}_{(J)}}(q^2)=f_{\mathrm{Dir}}\,e^{-aq^2}$ with free parameters $f_{\mathrm{Dir}}$ and $a$ \cite{Chen:2020xho}, which is an approximate description for the cross section of a direct production process in a narrow energy region between $\sqrt{s}=2.0\sim 2.6$ GeV. In addition, $m_{\mathcal{V}}$ and $\Gamma_{\mathcal{V}}$ are resonance parameters of the selected intermediate light vector meson states, which can be fixed by the corresponding experimental values or theoretical predictions. The total amplitude of $e^+e^- \to K^{(*/\prime)}_{(J)}\bar{K}^{(*/\prime)}_{(J)}$ can be written as the sum of a direct amplitude and different resonance contributions, i.e.,
\begin{eqnarray}
\mathcal{M}_{\mathrm{Total}}^{K^{(*/\prime)}_{(J)}\bar{K}^{(*/\prime)}_{(J)}}=\mathcal{M}_{\mathrm{Direct}}^{K^{(*/\prime)}_{(J)}\bar{K}^{(*/\prime)}_{(J)}}
+e^{i\theta_n}\sum_{\mathcal{V}_n}\mathcal{M}_{\mathcal{V}_n}^{K^{(*/\prime)}_{(J)}\bar{K}^{(*/\prime)}_{(J)}},
\end{eqnarray}
where $\mathcal{V}_n$ stands for the selected intermediate vector resonances, and $\theta_n$ is the phase angle between the direct annihilation amplitude and the intermediate light vector meson contribution. With the above total amplitude, the cross section of $e^+e^- \to K^{(*/\prime)}_{(J)}\bar{K}^{(*/\prime)}_{(J)}$ can be calculated directly by
\begin{eqnarray}
\sigma(e^+e^- \to K^{(*/\prime)}_{(J)}\bar{K}^{(*/\prime)}_{(J)})=\int\frac{1}{64\pi s}\frac{1}{\left| p_{3\mathrm{cm}}\right|^2}\overline{\left| \mathcal{M}_{\mathrm{Total}}^{K^{(*/\prime)}_{(J)}\bar{K}^{(*/\prime)}_{(J)}} \right|^2} dt, \nonumber \\
\end{eqnarray}
where $p_{3\mathrm{cm}}$ stands for the corresponding momentum of $p_3$ in the CM frame of a reaction.

Before depicting the cross section of our discussed open-strange processes, we need to select the suitable intermediate light vector meson states and investigate their resonance contributions, which can directly determine the coupling constants $g_{\mathcal{V} K^{(*/\prime)}_{(J)}K^{(*/\prime)}_{(J)}}$ and $f_{\mathcal{V}}$. In this work, we mainly concern with the CM energy region between 2.0 and 2.6 GeV. Consequently, how to obtain a  mass spectrum of light vector mesons above 2.0 GeV is a crucial step before carrying out a theoretical analysis of the open-strange processes $e^+e^- \to K^{(*/\prime)}_{(J)}\bar{K}^{(*/\prime)}_{(J)}$. In the next section, we present the mass spectrum of light vector mesons by an unquenched quark model and discuss their decay behaviors.

\section{ Mass spectrum and decay behaviors of light vector mesons }\label{sec3}

The mass spectrum of light vector mesons around 2.0 GeV has been studied by various potential models in the past several decades \cite{Godfrey:1985xj,Barnes:1996ff,Stanley:1980zm,Ebert:2009ub,Ebert:2005ha,Ishida:1986vn}. However, the results of the different models obviously differ from each other. Focusing on the excited strangeonium $\phi(3S)$ state that is considered to be a potential candidate of the $Y(2175)$, the analysis of the Regge trajectories indicates that its mass should be around 1.92 GeV \cite{Wang:2012wa}. However, Barnes {\it et al.} predicted  a corresponding mass as 2.05 GeV \cite{Barnes:2002mu}. In Ref. \cite{Ishida:1986vn}, the authors employed a covariant oscillator quark model with one gluon exchange effect, where the mass of $3^3S_1$ $s\bar{s}$ vector state is predicted to be 2.25 GeV. Therefore, obviously, there exists room  for theorists to develop the phenomenological model to obtain a relatively reliable spectroscopy description of light vector mesons above 2.0 GeV.

For describing the highly excited light-flavor hadronic states, the relativistic effect and unquenched correction should be considered in the calculation. In this work, we adopt an unquenched relativized potential model to study the mass spectrum of light vector mesons, which has been widely applied to study other meson families from a kaon sector to a heavy quarkonium sector \cite{Pang:2017dlw,Song:2015nia,Song:2015fha,Wang:2016krl,Wang:2019mhs,Wang:2020prx,Wang:2018rjg}. Here, we propose that the unquenched correction for the light vector meson family can be determined by other experimentally established $S$-wave and $D$-wave light meson families involving pion, $\rho_{2}$, and $\rho_{3}$, which could provide a relatively reliable scaling point for the energy region above 2.0 GeV in our theoretical model.  Furthermore, the Okubo-Zweig-Iizuka ($\mathrm{OZI}$)-allowed strong decay behaviors of these vector mesons can be obtained by the quark pair creation (QPC) model without $\beta$ parameter dependence involved in  
the simple harmonic oscillator (SHO) wave function necessary for calculating the transition matrix element, where we take  the corresponding numerical meson wave function directly from our unquenched relativized potential model as input.
For the convenience of the readers, we give a concise introduction to the method adopted in this work  in Appendices \ref{appendix-a} and  \ref{appendix-b}.

\begin{table}[t]
	\caption{The mass spectra of the observed $S$-wave and $D$-wave light mesons calculated by the unquenched relativized potential model.  Here, the comparison of the theoretical results and experimental data is given. All masses are in units of MeV.}\label{states}
	\begin{center}
		\renewcommand{\arraystretch}{1.2}
		\tabcolsep=1.2pt
\setlength{\tabcolsep}{0.8mm}
{
		\begin{tabular}{cccc}
			\toprule[1pt]\toprule[1pt]
Predicted states            & Mass (The.) & Observed states & Mass (Exp.)        \\
\midrule[1pt]

$\pi(1S)$         & 131       &    $\pi$        &  135 \cite{Zyla:2020zbs}       \\
$\pi(2S)$         & 1265      &    $\pi(1300)$         & $1300\pm100$ \cite{Zyla:2020zbs}\\
$\pi(3S)$         & 1759      &    $\pi(1800)$         & $1810^{+9}_{-11}$ \cite{Zyla:2020zbs}\\
$\pi(4S)$         & 2115      &    $\pi(2070)$        & $2070\pm35$ \cite{Anisovich:2001pn} \\
$\pi(5S)$         & 2372      &    $\pi(2360)$         & $2360\pm25$ \cite{Anisovich:2001pn} \\
$\rho(1S)$        & 775       &    $\rho(770)$        & 775 \cite{Zyla:2020zbs}       \\
$\rho(2S)$        & 1413      &    $\rho(1450)$         & $1465\pm25$ \cite{Zyla:2020zbs} \\
$\rho(3S)$        & 1862      &    $\rho(1900)$         & $1880\pm30$ \cite{Aubert:2006jq} \\
$\omega(1S)$      & 775       &    $\omega(782)$        & 783 \cite{Zyla:2020zbs}       \\
$\omega(2S)$      & 1413      &    $\omega(1420)$         & $1410\pm60$ \cite{Zyla:2020zbs} \\
$\pi_2(1D)$         & 1650      &    $\pi_2(1670)$         & $1670.6^{+2.9}_{-1.6}$ \cite{Zyla:2020zbs}\\
$\pi_2(2D)$         & 2003      &    $\pi_2(2005)$         & $1963.6^{+17}_{-27}$ \cite{Zyla:2020zbs}\\
$\pi_2(3D)$         & 2278      &    $\pi_2(2285)$         & $2285\pm20\pm25$ \cite{Anisovich:2010nh}\\
$\omega(1D)$      & 1633      &    $\omega(1650)$         & $1670\pm30$ \cite{Zyla:2020zbs} \\
$\rho(1D)$        & 1633      &    $\rho(1700)$         & $1720\pm20$ \cite{Zyla:2020zbs} \\
$\rho(2D)$        & 2003      &    $\rho(2000)$         & $2000\pm30$ \cite{Bugg:2004xu} \\
$\rho_2(2D)$        & 2007      &    $\rho_2(1940)$         & $1940\pm40$ \cite{Anisovich:2002su} \\
$\rho_2(3D)$        & 2284      &    $\rho_2(2225)$         & $2225\pm35$ \cite{Anisovich:2002su} \\
$\rho_3(1D)$        & 1672      &    $\rho_3(1690)$         & $1688.8\pm2.1$ \cite{Zyla:2020zbs} \\
$\rho_3(2D)$        & 2013      &    $\rho_3(1990)$         & $1982\pm14$ \cite{Anisovich:2002su} \\
$\rho_3(3D)$        & 2284      &    $\rho_3(2250)$         & $2290\pm20\pm30$ \cite{Amelin:2000nm} \\
$\phi(1S)$        & 1020      &    $\phi(1020)$         & 1019 \cite{Zyla:2020zbs}       \\
$\phi(2S)$        & 1665      &    $\phi(1680)$         & $1680\pm20$ \cite{Zyla:2020zbs} \\
       \bottomrule[1pt]\bottomrule[1pt]
		\end{tabular}
}
	\end{center}
\end{table}

\begin{table}
\caption{ Parameters in the unquenched relativized potential model adopted in this work.}
\renewcommand\arraystretch{1.2}
\begin{tabular*}{80mm}{c@{\extracolsep{\fill}}cccc}
\toprule[1pt]
Parameter                 & Value  & Parameter                 & Value  \\
\toprule[0.8pt]
$b$ (GeV$^2$)             & 0.229  & $c$ (GeV)                 & -0.300  \\
$\epsilon_{\mathrm{sos}}$ & 0.973  & $\epsilon_c$              & -0.164  \\
$\epsilon_{\mathrm{sov}}$ & 0.262  & $\epsilon_t$              & 1.993  \\
$\mu$ (GeV)               & 0.081  &                           &        \\
\toprule[1pt]
\end{tabular*}\label{parameter}
\end{table}

\begin{table}
\caption{A comparison of the decay behavior of some well-established light vector mesons of the experimental average values and our theoretical estimate. Here, $\mathcal{B}_{e^+e^-}(\mathcal{V})$ and $\mathcal{B}_{h_1h_2}(\mathcal{V})$ represent the branching ratio $\Gamma(\mathcal{V} \to e^+e^-)/\Gamma^{Total}_{\mathcal{V}}$ and $\Gamma(\mathcal{V} \to h_1h_2)/\Gamma^{Total}_{\mathcal{V}}$, respectively. }
\renewcommand\arraystretch{1.2}
\begin{tabular*}{86mm}{c@{\extracolsep{\fill}}cccc}
\toprule[1pt]
Decay       &  Our  &  Exp. (Ave.)  \cite{Zyla:2020zbs}  \\
\toprule[0.8pt]
$\Gamma_{e^+e^-}(\rho(770)  )$ ~~~~(keV)           & 6.98                & $7.04\pm0.06$  \\
$\mathcal{B}_{e^+e^-}(\rho(1450) )\mathcal{B}_{\omega\pi}(\rho(1450) )$ ($\times10^{-6}$)          & 4.4                & $3.7\pm0.4$  \\
$\mathcal{B}_{e^+e^-}(\rho(1450) )\mathcal{B}_{\rho\eta}(\rho(1450) )$ ($\times10^{-6}$)          & 0.60                & $0.58\pm0.07$  \\
$\Gamma_{e^+e^-}(\omega(782)  )$~~~~ (keV)           & 0.78                & $0.60\pm0.02$  \\
$\mathcal{B}_{e^+e^-}(\omega(1420) )\mathcal{B}_{\omega\eta}(\omega(1420) )$ ($\times10^{-7}$)          & 0.11                & $0.29\pm0.15$  \\
$\mathcal{B}_{e^+e^-}(\omega(1420) )\mathcal{B}_{\rho\pi}(\omega(1420) )$ ($\times10^{-7}$)          & 2.74                & $6.58\pm1.49$  \\
$\Gamma_{e^+e^-}(\phi(1020) )$~~~~ (keV)           & 3.19               & $1.27\pm0.04$  \\
$\mathcal{B}_{e^+e^-}(\phi(1680) )\mathcal{B}_{KK^*}(\phi(1680))$ ($\times10^{-7}$)          & 20.4                & $22.2\pm8.7$  \\
$\mathcal{B}_{e^+e^-}(\phi(1680) )\mathcal{B}_{K^0\bar{K}^0}(\phi(1680) )$ ($\times10^{-7}$)          & 1.37                & $1.31\pm0.59$  \\

\toprule[1pt]
\end{tabular*}\label{comparison}
\end{table}


As pointed out in Ref. \cite{Wang:2018rjg}, the color screened effect is mainly reflected on the highly excited hadronic states and so the key point of obtaining the reliable mass spectrum of light vector mesons above 2.0 GeV is to determine the screened parameter $\mu$ in the unquenched relativized potential model. In fact, the unquenched effect from the color screened interaction is a kind of nonperturbative behavior of QCD and is difficult to be solved from a theoretical perspective, so its quantification needs the input of available experimental hints. At present, within the messy research situation of the light-flavor vector mesons, we believe that other light meson families could provide some valuable hints on our theoretical model.

In the quark model, in addition to an $S$-wave vector meson state, there still exist the corresponding vector $D$-wave partners, i.e., $n^{2S+1}L_J=n^3D_1$. Thus, as their spin singlet or triplet states, the experimental information of an $S$-wave pseudoscalar pion family and $D$-wave $\pi_2$, $\rho_2$, and $\rho_3$ states can help us determine the potential model parameters, where we notice that the masses of some highly excited states can reach around 2.3 GeV, such as $\pi(2360)$ \cite{Anisovich:2001pn}, $\pi_2(2285)$ \cite{Anisovich:2010nh}, $\rho_2(2225)$ \cite{Anisovich:2002su} and $\rho_3(2250)$ \cite{Amelin:2000nm}. Thus, these highly excited states play the role of the important scaling point of color screened effect. The measured masses of these light-flavor mesons are summarized in Table \ref{states}, which can be used to constrain three main screened confinement parameters and four relativistic correction factors $\epsilon_i$. Based on a set of parameters listed in Table \ref{parameter}, the global aspect of the light-flavor $S$-wave and $D$-wave meson mass spectra is well consistent with  those of experiments, which can be clearly shown by the comparison between experimental values and our theoretical estimates in Table \ref{states}. The screening parameter $\mu=0.81$ indicates that the unquenched effect is indeed significant for describing the mass spectrum of light vector mesons.

With the above preparation, we can directly predict the mass spectrum and decay behavior of higher $\rho$, $\omega$, and $\phi$ mesonic states above 2.0 GeV. In the QPC model, since the meson wave functions of initial and final states can be fixed by the corresponding eigenvectors solved from the above unquenched potential model, the partial width of each decay channel is only dependent on the parameter $\gamma$. However, it is worth noting that the corresponding branching ratio is independent on any parameter because the $\gamma$ in the numerator and the denominator cancel out each other. In addition, the di-lepton widths of the higher vector states can also be related to the zero-point behavior of their radial wave functions, whose concrete width formula can be found in Refs. \cite{Wang:2020kte,Godfrey:1985xj}. Thus, the production rates of intermediate light vector mesons in $e^+e^-$ annihilations can also be estimated from a theoretical perspective. In Table \ref{comparison}, we show the decay behavior comparison of some well-established light vector mesons  between the experimental values and our theoretical results, which involve $\rho(770)$, $\rho(1450)$, $\omega(782)$, $\omega(1420)$, $\phi(1020)$, and $\phi(1680)$. It can be seen that our theoretical predictions are well consistent with measured results, in a sense, which can prove the reliability  of our models in estimating the decay properties of higher light vector meson states above 2.0 GeV.

The calculated mass spectrum and branching ratios of two-body open-strange strong decays as well as di-lepton decay widths of $\rho$($\omega$) and $\phi$ meson states above 2.0 GeV are listed in Tables \ref{rhoomega} and \ref{phi}, respectively. There are eight  light vector mesons existing near the energy region of $Y(2175)$, which include $\rho(2D)/\omega(2D)$, $\rho(4S)/\omega(4S)$, $\rho(3D)/\omega(3D)$, $\phi(3S)$ and $\phi(2D)$. However, it is almost impossible to study the present experimental data of open-strange channels by adding so many resonance contributions. Thus, we have to select the main intermediate resonances in our practical analysis.  From the calculated numerical results, the di-lepton widths of the isoscalar $\omega$ states are one order smaller than those of the isovector $\rho$ partners, and their branching ratios to open-strange decay channels are almost the same. Furthermore, one can see that the branching ratios of $\phi(3S)/\phi(2D) \to K^{(*/\prime)}_{(J)}\bar{K}^{(*/\prime)}_{(J)}$ are far larger than those of $\rho(2D)$, $\rho(4S)$, and $\rho(3D)$ although there is no obvious difference among their production rates via electron-positron collisions. Therefore, we can conclude that the resonance contributions from the highly excited $\rho$ and $\omega$ states can be safely ignored in studying $e^+e^- \to K^{(*/\prime)}_{(J)}\bar{K}^{(*/\prime)}_{(J)}$ reaction above 2.0 GeV. 
Additionally, it is worth mentioning that the recently observed structure near 2.2 GeV in $e^+e^- \to K^+K^-$ by BESIII has been collected into $``\rho(2170)"$ in the 2020 version of PDG \cite{Zyla:2020zbs}, which should need a careful judgement and is not supported by our theoretical results here. 

In Tables \ref{rhoomega} and \ref{phi},  we also give the theoretical total widths of the light vector mesons by taking a typical value of $\gamma=6.57$ \cite{Pang:2019ttv}, which can be used as a reference. One can see that $\phi$ states play an absolutely dominant role in the open-strange processes from $e^+e^-$ collisions, so three strangeonium states $\phi(3S)$, $\phi(2D)$, and $\phi(4S)$ will be considered in our analysis. Utilizing the calculated decay behaviors of the higher strange quarkonium states presented in Table \ref{phi}, in the next section, we can directly decipher the cross sections of the present reported open-strange processes based on $e^+e^-$ collisions.



\begin{table*}
\caption{The branching ratios of two-body open-strange strong decay and di-lepton widths of $\rho$ and $\omega$ meson states above 2.0 GeV. Here, the tiny branching ratio is marked as ``$\cdots$". }
\renewcommand{\arraystretch}{1.2}
\begin{tabular*}{175mm}{c@{\extracolsep{\fill}}ccccccccccc}
\toprule[1pt]\toprule[1pt]
States                           & $\rho(2D)$ & $\rho(4S)$ & $\rho(3D)$ & $\rho(5S)$  & $\omega(2D)$& $\omega(4S)$& $\omega(3D)$& $\omega(5S)$ \\
\toprule[0.8pt]
Mass (GeV)                       & 2.003      & 2.180      & 2.283      & 2.422 & 2.003       & 2.180       & 2.283       & 2.422       \\
$\Gamma_{e^+e^-}$ (keV)          & 0.020      & 0.063      & 0.016      & 0.036    & 0.0022      & 0.007       & 0.0018      & 0.004 \\
$\Gamma_{\mathrm{Total}}$ (MeV)  & 179      & 102      & 158      & 80   & 181       & 104       & 94        & 69   \\
$\mathcal{B}(KK)$                 & 0.006      & 0.002      & 0.002      & $\cdots$ & 0.006       & 0.002       & 0.003       & $\cdots$  \\
$\mathcal{B}(KK^*)$               & 0.001      & $\cdots$   & $\cdots$   & $\cdots$ & 0.001       & $\cdots$    & $\cdots$    & $\cdots$ \\
$\mathcal{B}(K^*K^*)$             & $\cdots$   & 0.003      & $\cdots$   & 0.003  & $\cdots$    & 0.003       & 0.001       & 0.003      \\
$\mathcal{B}(KK_1(1270))$         & $\cdots$   & 0.010      & $\cdots$   & 0.004  & $\cdots$    & 0.010       & $\cdots$    & 0.005    \\
$\mathcal{B}(KK^*(1410))$         & $\cdots$      & 0.001      & $\cdots$      & $\cdots$  & $\cdots$       & 0.001       & $\cdots$      & $\cdots$      \\
$\mathcal{B}(K^*K_1(1270))$       & $\cdots$   & 0.004      & $\cdots$   & 0.004  & $\cdots$    & 0.004       & $\cdots$    & 0.005      \\
$\mathcal{B}(KK(1460))$           & 0.002      & 0.003      & 0.003      & 0.001  & 0.002       & 0.002       & 0.005       & 0.001      \\
\toprule[1pt]\toprule[1pt]
\end{tabular*}\label{rhoomega}
\end{table*}

\begin{table*}
\caption{The branching ratios of two-body open-strange strong decay and di-lepton widths of $\phi$ mesons above 2.0 GeV. Additionally, the mass and decay behaviors of missing $\phi(1D)$ in experiments are also given. Here, the tiny branching ratio is marked as ``$\cdots$".}
\renewcommand{\arraystretch}{1.2}
\begin{tabular*}{170mm}{c@{\extracolsep{\fill}}cccccc}
\toprule[1pt]\toprule[1pt]
States                               & $\phi(1D)$  & $\phi(3S)$  & $\phi(2D)$  & $\phi(4S)$  & $\phi(3D)$ \\
\toprule[0.8pt]
Mass (GeV)             & 1.860       & 2.103       & 2.236       & 2.423       & 2.519      \\
$\Gamma_{e^+e^-}$ (keV)              & 0.063       & 0.106       & 0.017       & 0.050       & 0.010      \\
$\Gamma_{\mathrm{Total}}$ (MeV)      & 515       & 156       & 265       & 140       & 171      \\
$\mathcal{B}(KK)$                    & 0.076       & 0.059       & 0.087       & 0.047       & 0.084      \\
$\mathcal{B}(KK^*)$                  & 0.100       & 0.242       & 0.067       & 0.145       & 0.050      \\
$\mathcal{B}(K^*K^*)$                & 0.016       & 0.007       & 0.105       & 0.014       & 0.139      \\
$\mathcal{B}(KK^*(1410))$            & $\cdots$    & 0.180       & 0.080       & 0.142       & 0.052      \\
$\mathcal{B}(KK^*(1680))$            & $\cdots$    & $\cdots$    & $\cdots$    & 0.004       & 0.003      \\
$\mathcal{B}(K^*K^*(1410))$          & $\cdots$    & $\cdots$    & $\cdots$    & $\cdots$       & 0.045      \\
$\mathcal{B}(KK_2^*(1430))$          & $\cdots$    & 0.115       & 0.043       & 0.091       & 0.043      \\
$\mathcal{B}(K^*K_2^*(1430))$        & $\cdots$    & $\cdots$    & $\cdots$    & 0.057       & 0.047      \\
$\mathcal{B}(KK_1(1270))$            & 0.799       & 0.185       & 0.356       & 0.154       & 0.293      \\
$\mathcal{B}(K^*K_1(1270))$          & $\cdots$    & $\cdots$    & 0.110       & 0.018       & 0.095      \\
$\mathcal{B}(KK_1(1400))$            & $\cdots$    & 0.064       & 0.026       & 0.027       & 0.015      \\
$\mathcal{B}(K^*K_1(1400))$          & $\cdots$    & $\cdots$    & $\cdots$    & 0.126       & 0.003      \\
$\mathcal{B}(KK(1460))$              & $\cdots$    & 0.127       & 0.112       & 0.073       & 0.101      \\
$\mathcal{B}(K^*K(1460))$            & $\cdots$    & $\cdots$    & $\cdots$    & 0.013       & 0.005      \\
$\mathcal{B}(K^*K_0(1430))$          & $\cdots$    & $\cdots$    & $\cdots$    & 0.049       & $\cdots$   \\
$\mathcal{B}(KK_3^*(1780))$          & $\cdots$    & $\cdots$    & $\cdots$    & 0.021       & 0.014      \\

\toprule[1pt]\toprule[1pt]
\end{tabular*}\label{phi}
\end{table*}

\section{ results of numerical analysis}\label{sec4}

\begin{figure*}[t]
	\includegraphics[width=14.0cm,keepaspectratio]{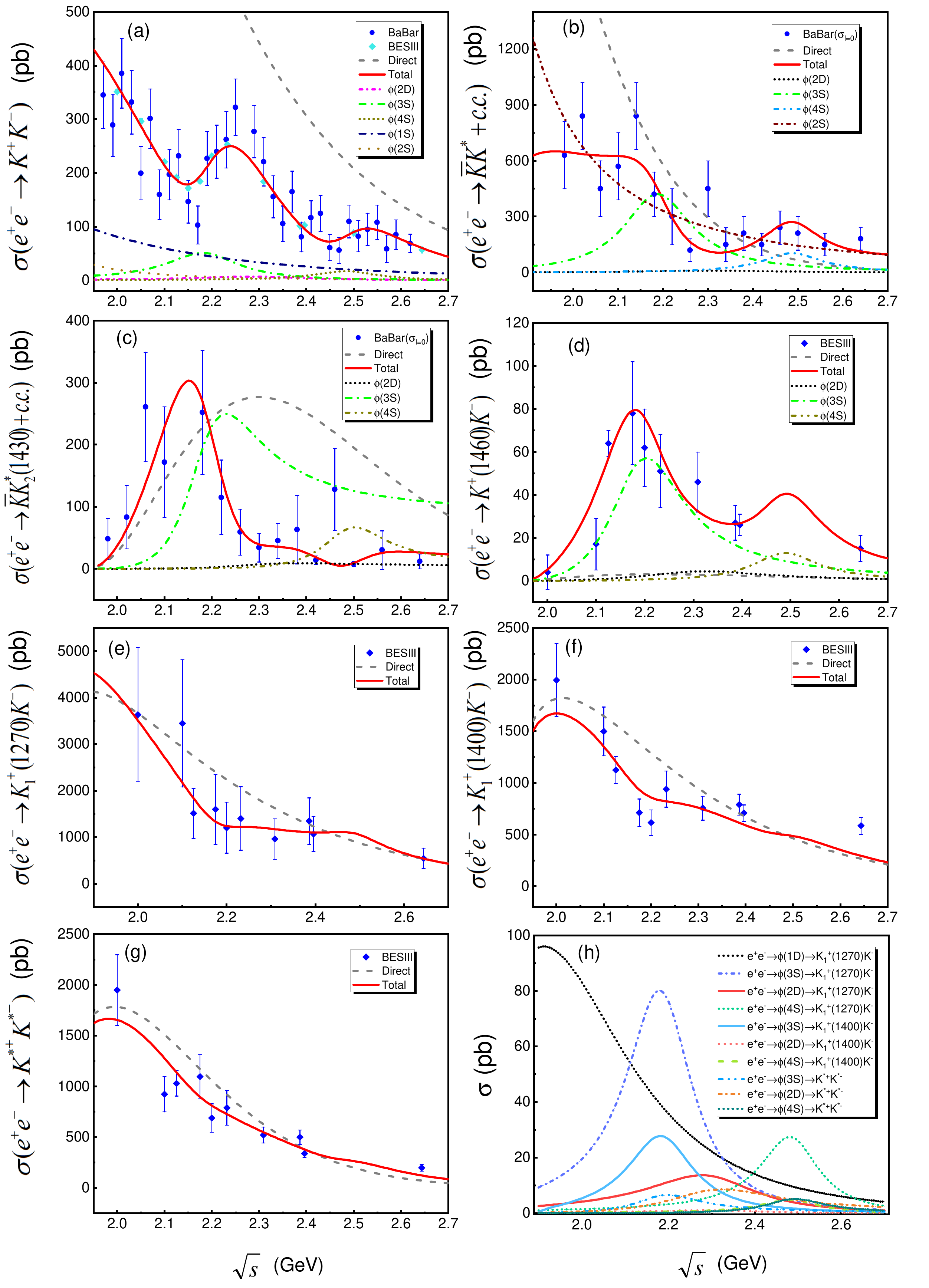}
	\caption{ The combined analysis to seven open-strange processes from $e^+e^-$ collisions \cite{Ablikim:2018iyx,Ablikim:2020pgw,BaBar:2013jqz,Aubert:2007ym}, which are shown in figure (a)-(g), successively. Here, $\sigma_{I=0}$ in figure (b) and (c) means that the measured cross section of reaction process corresponds to the isoscalar component. }\label{Y2175analysis}
\end{figure*}

In the following, we fit the cross sections for seven open-strange reactions of $e^+e^- \to K^+{K}^-$ \cite{Ablikim:2018iyx,BaBar:2013jqz}, $e^+e^- \to K\bar{K}^{*}+c.c.$ \cite{Aubert:2007ym}, $e^+e^- \to K^{*+}K^{*-}$ \cite{Ablikim:2020pgw}, $e^+e^- \to K_1(1270)^+K^-$ \cite{Ablikim:2020pgw}, $e^+e^- \to K_1(1400)^+K^-$ \cite{Ablikim:2020pgw}, $e^+e^- \to K_2^{*}(1430)\bar{K}+c.c.$ \cite{Aubert:2007ym} and $e^+e^- \to K(1460)^+K^-$ \cite{Ablikim:2020pgw}, which provide us a direct evidence to  demonstrate the nature of the vector structure around 2.2 GeV observed in these processes. It is worth mentioning that although the contributions from resonances $\phi(1020)$, $\phi(1680)$, and $\phi(1D)$ in the energy region above 2.0 GeV are obviously suppressed compared with those at their resonant peak positions, they still may play a significant background role in some specific open-strange channels, which have to be included in our analysis. According to the corresponding decay behaviors presented in Tables \ref{comparison} and \ref{phi}, we consider the contributions of lower $\phi(1020)$, $\phi(1680)$, and $\phi(1D)$ in reaction channels $KK$, $KK^{(*)}$, and $KK_1(1270)$, respectively. The resonance masses and total widths of $\phi(1020)$ and $\phi(1680)$ can be fixed by the experimental average values in PDG \cite{Zyla:2020zbs}.  All coupling constants $g_{\phi K^{(*/\prime)}_{(J)}K^{(*/\prime)}_{(J)}}$/$f_{\phi}$ in Eqs. (\ref{eq22})-(\ref{eq26}) can be related to the corresponding products of $\Gamma(\phi \to e^+e^-)\times \mathcal{B}(\phi \to K^{(*/\prime)}_{(J)}\bar{K}^{(*/\prime)}_{(J)})$ by the following expressions
\begin{eqnarray}
\frac{g_{\phi KK}}{f_{\phi}}&=&\sqrt{\frac{288\pi^2\Gamma_{\phi}m_{\phi}\Gamma(\phi \to e^+e^-)\mathcal{B}(\phi \to K\bar{K})}{e^4\lambda(m_{\phi},m_K,m_K)^{\frac{3}{2}}}}, \nonumber \\
\frac{g_{\phi KK^{*}}}{f_{\phi}}&=&\sqrt{\frac{144\pi^2\Gamma_{\phi}\Gamma(\phi \to e^+e^-)\mathcal{B}(\phi \to K\bar{K}^{*}+c.c.)}{e^4\lambda(m_{\phi},m_K,m_{K^{*}})^{\frac{3}{2}}m_{\phi}}}, \nonumber \\
\frac{g_{\phi K^*K^{*}}}{f_{\phi}}&=&\sqrt{\frac{288\pi^2\Gamma_{\phi}m_{\phi}\Gamma(\phi \to e^+e^-)\mathcal{B}(\phi \to K^*\bar{K}^{*})}{e^4\lambda(m_{\phi},m_{K^{*}},m_{K^{*}})^{\frac{3}{2}}(m_{\phi}^4-4m_{\phi}^2m_{K^{*}}^2+12m_{K^{*}}^4)}}, \nonumber \\
\frac{g_{\phi KK_1}}{f_{\phi}}&=&\sqrt{\frac{144\pi^2\Gamma_{\phi}m_{\phi}\Gamma(\phi \to e^+e^-)\mathcal{B}(\phi \to K\bar{K}_1+c.c.)}{e^4\lambda(m_{\phi},m_K,m_{K_1})^{\frac{1}{2}}}}, \nonumber \\
\frac{g_{\phi KK_2}}{f_{\phi}}&=&\sqrt{\frac{144\pi^2\Gamma_{\phi}\Gamma(\phi \to e^+e^-)\mathcal{B}(\phi \to K\bar{K_2}+c.c.)}{e^4\lambda(m_{\phi},m_K,m_{K_2})^{\frac{5}{2}}m_{\phi}^3m_{\phi}^2}}, \nonumber \\
\frac{g_{\phi KK^{\prime}}}{f_{\phi}}&=&\sqrt{\frac{144\pi^2\Gamma_{\phi}m_{\phi}\Gamma(\phi \to e^+e^-)\mathcal{B}(\phi \to K\bar{K}^{\prime}+c.c.)}{e^4\lambda(m_{\phi},m_K,m_K^{\prime})^{\frac{3}{2}}}}, \nonumber \\
\end{eqnarray}
where $\lambda(x,y,z)=x^2+y^2+z^2-2xy-2xz-2yz$ is the K$\ddot{\mathrm{a}}$llen function. Thus, the only fitting parameters are relative phase angles and $f_{\mathrm{Dir}}$ and $a$ in a direct production amplitude. However, in a practical analysis, we find that the fitted $\chi^2$ value is obviously dependent on resonance parameters of three strange quarkonium states $\phi(3S)$, $\phi(2D)$, and $\phi(4S)$. At the same time, combining the fact that the measured errors of experimental resonance parameters of most of light-flavor mesons are generally large as seen in Table \ref{states}, the masses of $m_{\phi(3S)}$, $m_{\phi(2D)}$, $m_{\phi(4S)}$, and strong decay parameter $\gamma$ can be set as free parameters, which should vary around their fixed theoretical values.

The combined fit to seven open-strange processes from $e^+e^-$ collisions mentioned above is presented in Fig. \ref{Y2175analysis}. Benefiting from the calculated decay behaviors of light vector mesons in Sec. \ref{sec3}, the contributions of each individual light vector meson to the cross sections of different open-strange processes can be directly obtained, which are also shown in Fig. \ref{Y2175analysis}. One can see that experimental cross sections of the measured open-strange processes can be described well by introducing the resonance contributions from highly excited $3S$, $2D$, and $4S$ strange quarkonium states and a combined $\chi^2/d.o.f=2.29$ is obtained. Especially, the apparent enhancement or dip around 2.2 GeV seen in $e^+e^- \to K^+K^-, K(1460)^+K^-, K_1(1270)^+K^-$, $K_1(1400)^+K^-$, and $K_2^{*}(1430)\bar{K}+c.c.$ can be completely reproduced via the interference effect from the resonance contributions of $\phi(3S)$ and $\phi(2D)$ and the continuum background, where an $S$-wave state plays a dominant role. Here, it is worth mentioning that the $e^+e^- \to K^*\bar{K}_1(1270)+c.c.$ could be an excellent reaction process to detect the dominant signal of a $D$-wave state, where the contribution from a $3S$ state is suppressed as shown in Table \ref{phi}. Based on the interference effect, we can naturally explain why there exists an inconsistent peak position of the observed vector structure around 2.2 GeV among different reaction channels, such as $KK$ channel in Fig. \ref{Y2175analysis}(a) and other channels in Figs. \ref{Y2175analysis}(c)-(f). Here, we can see that a simple Breit-Wigner fit to the corresponding cross section is indeed a very rough treatment.

The obtained phase angles and direct production parameters of each process in the combined fit are summarized in Table \ref{para}. In addition, the resonance masses and widths of $\phi(3S)$, $\phi(2D)$, and $\phi(4S)$ are fitted to be
\begin{eqnarray}
m_{\phi(3S)}=2183 \pm 1 ~\mathrm{MeV}, \Gamma_{\phi(3S)}=185 \pm4 ~\mathrm{MeV}, \nonumber
\end{eqnarray}
\begin{eqnarray}
m_{\phi(2D)}=2290\pm3 ~\mathrm{MeV}, \Gamma_{\phi(2D)}=312 \pm 6 ~\mathrm{MeV}, \nonumber\\
m_{\phi(4S)}=2485\pm5 ~\mathrm{MeV}, \Gamma_{\phi(4S)}=165 \pm 3 ~\mathrm{MeV}, \nonumber
\end{eqnarray}
respectively. The above results indicate that both $\phi(3S)$ and $\phi(2D)$ are broad states, which are consistent with previous theoretical calculations \cite{Barnes:2002mu,Pang:2019ttv,Wang:2012wa,Ding:2007pc}. As mentioned in Introduction, the measured resonance width of $Y(2175)$ is narrower than predictions for $\phi(3S)$ or $\phi(2D)$.
From Fig. \ref{Y2175analysis}, it can be seen that this width problem can also be explained in our theoretical analysis of considering the interference effect among two light vector meson contributions as well as a background term. Here, if we fit the cross sections in Fig. \ref{Y2175analysis}(a) or \ref{Y2175analysis}(c)-(f) by a simple Breit-Wigner formula, we would obtain the narrow width around 100 MeV.

Additionally, we notice that the experimental data of reaction $e^+e^- \to K^{*+}K^{*-}$ in Fig. \ref{Y2175analysis}(g) does not show an obvious signal for the existence of a vector structure around 2.2 GeV. Here, we want to emphasize that this interesting phenomenon can also be understood in our theoretical framework. From Table \ref{phi}, we can find a branching ratio $\mathcal{B}(\phi(3S)\to K^*\bar{K}^*)=7\times 10^{-3}$, which is obviously smaller than those of other open-strange decay channels. The reason for this is that there exists a node effect for the overlap integral among the wave functions of initial $\phi(3S)$ state and final state $K^*\bar{K}^*$. From Fig. \ref{Y2175analysis}(h), it can be seen that these resonance contributions in process $e^+e^- \to K^{*+}K^{*-}$ are of the order of several pb, which is far lower than a direct continuum.  Therefore, this is a strong evidence to support our explanation to the vector structure around 2.2 GeV observed in open-strange processes. In conclusion, our theoretical analysis can well clarify the present puzzling situation in understanding the cross section of open-strange processes from $e^+e^-$ annihilation without adding any unknown resonance contributions, which means that this work provides a new opinion to construct the light vector meson family and understand the $Y(2175)$.

Finally, we would like to suggest our experimental colleagues to search for a predicted new resonance structure $\phi(4S)$ around 2.5 GeV, whose signal can be discovered by several typical open-strange processes $e^+e^- \to K\bar{K}$, $e^+e^- \to K\bar{K}^*+c.c.$, and $e^+e^- \to K\bar{K}(1460)+c.c.$ as presented in Fig. \ref{Y2175analysis}, especially for $KK(1460)$ channel, where the experimental data on relevant energy region is still lacking. This should be an interesting research topic on the future BESIII and BelleII experiments.

\begin{table*}
\caption{ The fitted parameters in the combined analysis to the experimental data of seven open-strange processes from $e^+e^-$ collisions. The listed phase angles $\theta_{\phi}$ are in units of radian. }
\renewcommand\arraystretch{1.2}
\begin{tabular*}{175mm}{c@{\extracolsep{\fill}}ccccccccc}
\toprule[1pt]\toprule[1pt]
Processes & $f_{\mathrm{Dir}}$ &  $a$ ($\mathrm{GeV}^{-2}$) &  $\theta_{\phi(1S)}$ &  $\theta_{\phi(2S)}$ &  $\theta_{\phi(1D)}$ & $\theta_{\phi(3S)}$ & $\theta_{\phi(2D)}$ &  $\theta_{\phi(4S)}$ &          \\
\toprule[0.8pt]
$e^+e^- \to K^+{K}^-$        & $-0.29$  & $0.33$ & $6.09\pm0.04$ & $0.52\pm0.06$ & $\cdots$  & $2.54\pm0.02$ & $5.77\pm0.05$ & $2.80\pm0.05$ \\
$e^+e^- \to K\bar{K}^{*}+c.c.$   & $1.95\pm0.07$  & $0.85\pm0.01$ & $\cdots$ & $3.51\pm0.10$ & $\cdots$  & $2.74\pm0.17$ & $1.32\pm0.41$ & $4.84\pm0.38$ \\
$e^+e^- \to K^{*+}{K}^{*-}$   & $-2.68\pm0.05$  & $0.89\pm0.01$ & $\cdots$ & $\cdots$ & $\cdots$  & $2.14\pm0.47$ & $2.57\pm0.36$ & $3.05\pm0.49$ \\
$e^+e^- \to K_1(1270)^{+}{K}^{-}$   & $0.78\pm0.07$  & $0.22\pm0.01$ & $\cdots$ & $\cdots$ & $2.47\pm0.52$  & $5.02\pm0.63$ & $4.87\pm1.10$ & $1.28\pm0.95$ \\
$e^+e^- \to K_1(1400)^{+}{K}^{-}$   & $0.87\pm0.16$  & $0.28\pm0.04$ & $\cdots$ & $\cdots$ & $\cdots$  & $4.62\pm0.21$ & $5.79\pm0.65$ & $6.28\pm0.33$ \\
$e^+e^- \to K_2^{*}(1430)\bar{K}+c.c.$  & $1.09\pm0.11$  & $0.80\pm0.02$ & $\cdots$ & $\cdots$ & $\cdots$  & $3.54\pm0.10$ & $1.11\pm0.51$ & $4.67\pm0.24$ \\
$e^+e^- \to K(1460)^{+}{K}^{-}$      & $-0.12\pm0.22$  & $0.46\pm0.35$ & $\cdots$ & $\cdots$ & $\cdots$  & $6.22\pm0.18$ & $5.55\pm0.64$ & $6.26\pm1.30$ \\
\toprule[1pt]\toprule[1pt]
\end{tabular*}\label{para}
\end{table*}

\section{Summary}\label{sec5}

Recently, the BESIII Collaboration performed the precise measurements of cross sections for open-strange processes $e^+e^- \to K^+K^-$ \cite{Ablikim:2018iyx} and $e^+e^- \to K^+K^-\pi^0 \pi^0$ \cite{Ablikim:2020pgw}, where a clear structure around 2.2 GeV was observed. This is the first certain evidence for the existence of the vector structure around 2.2 GeV in open-strange reaction processes. The experimental resonance masses and widths of this newly observed vector structure are found to be consistent with those of the $Y(2175)$ firstly reported in hidden-strange reaction $e^+e^- \to \phi f_0(980) \to \phi \pi^+\pi^-$ by the BaBar Collaboration in 2006 \cite{Aubert:2006bu}. Thus, this simple comparison seems to imply that this structure around 2.2 GeV may be the same state as $Y(2175)$. In the past years, the popular theoretical explanations of $Y(2175)$ mainly include hybrid $s\bar{s}g$ \cite{Ding:2006ya}, vector strangeonium state $\phi(3S)$ \cite{Barnes:2002mu,Pang:2019ttv}, and $\phi(2D)$ \cite{Wang:2012wa,Ding:2007pc,Pang:2019ttv}, which are just favored by the open-strange channels.  However, the theoretical studies on their decay behaviors \cite{Ding:2006ya,Barnes:2002mu,Pang:2019ttv,Wang:2012wa,Ding:2007pc,Close:1994hc,Close:2003mb,Zhu:1998sv,Zhu:1999wg} definitely indicate that it is difficult to understand the experimental data of these open-strange reactions \cite{Ablikim:2018iyx,Ablikim:2020pgw} either under the hybrid or strangeonium assignment to the $Y(2175)$. This means that there are no appropriate theoretical pictures to convincingly explain this newly observed vector structure around 2.2 GeV.

In order to clarify this puzzling situation, we have pointed out that it is not an easy task to analyze the cross sections of the open-strange processes from $e^+e^-$ annihilations, where both highly excited and light vector mesons $\rho$, $\omega$, and $\phi$ states may have significant contributions.  Thus, a simple Breit-Wigner fit to the signal around 2.2 GeV observed in the cross sections of open-strange processes must be a very rough treatment. Based on this motivation, in this work, we have performed a combined analysis to the cross sections of seven reported open-strange reactions of $e^+e^- \to K^+{K}^-$ \cite{Ablikim:2018iyx,BaBar:2013jqz}, $e^+e^- \to K\bar{K}^{*}+c.c.$ \cite{Aubert:2007ym}, $e^+e^- \to K^{*+}K^{*-}$ \cite{Ablikim:2020pgw}, $e^+e^- \to K_1(1270)^+K^-$ \cite{Ablikim:2020pgw}, $e^+e^- \to K_1(1400)^+K^-$ \cite{Ablikim:2020pgw}, $e^+e^- \to K_2^{*}(1430)\bar{K}+c.c.$ \cite{Aubert:2007ym} and $e^+e^- \to K(1460)^+K^-$ \cite{Ablikim:2020pgw} by introducing the light vector meson contributions, which is supported by the study of hadron spectroscopy. Here, we have employed an unquenched relativized potential model \cite{Pang:2017dlw,Song:2015nia,Song:2015fha,Wang:2016krl,Wang:2019mhs,Wang:2020prx,Wang:2018rjg} to study the mass spectra and wave functions of light vector meson states. By taking the exact numerical wave functions from the potential model as input, the di-lepton widths and the branching ratios of open-strange decay channels of light vector mesons can be estimated without any parameter dependence. Based on these available theoretical results, the cross sections of the seven open-strange reactions are found to be well described by considering the direct production and  the light vector meson contributions. Furthermore, we have demonstrated that the observed vector structure around 2.2 GeV in these open-strange channels does not correspond to a single resonance and it can be identified as an interference signal from highly excited strange quarkonium states $\phi(3S)$ and $\phi(2D)$, by which the present puzzling situations can be naturally clarified.  From this practical example in this work, we have actually  provided a new perspective to construct the light vector meson family and understand the $Y(2175)$, which can continue to be tested in hidden-strange channels and other related processes. These should be helpful to solve the messy research situation of light vector mesons around 2.2 GeV thoroughly in the future.




In the full hadron spectrum, the light meson spectroscopy has a particular position because there exist abundant experimental data in the energy region of light mesons. In addition to the intriguing vector structure around 2.2 GeV, we have also predicted a new strange quarkonium structure $\phi(4S)$ near $\sqrt{s}=2.5$ GeV, whose mass and total width are predicted to be $m_{\phi(4S)}=2485\pm5 ~\mathrm{MeV}$ and $ \Gamma_{\phi(4S)}=165 \pm 3 ~\mathrm{MeV}$, respectively. The establishment of this particle in experiments is very important for further constructing the light vector meson family. This should be a challenging task to the experimentalist community and is worth working in the future.

\appendix

\section{A concise review of the unquenched relativized potential model}\label{appendix-a}

The Hamiltonian depicting the interaction between quark and antiquark in the unquenched relativized potential model is described as \cite{Pang:2017dlw,Song:2015nia,Song:2015fha,Wang:2016krl,Wang:2019mhs,Wang:2020prx,Wang:2018rjg,Godfrey:1985xj}
\begin{eqnarray}
\widetilde{H}=\widetilde{H}_0+\widetilde{V}_{\mathrm{eff}}(\mathbf{p},\mathbf{r}) \label{eq1}
\end{eqnarray}
with
\begin{eqnarray}
\widetilde{H}_0=\sqrt{m_q^2+\mathrm{p^2}}+\sqrt{m_{\bar{q}}^2+\mathrm{p^2}},
\end{eqnarray}
where $m_q$ and $m_{\bar{q}}$ denote the constituent masses of light quark and antiquark, respectively. In our calculation, the quark masses $m_{u}=m_{d}=
0.22$ GeV and $m_s=0.424$ GeV are taken. The effective potential of $q\bar{q}$ interaction contains three parts, i.e.,
\begin{eqnarray}
\widetilde{V}_{\mathrm{eff}}(\mathbf{p},\mathbf{r})=\widetilde{H}^{\mathrm{conf}}+\widetilde{H}^{\mathrm{so}}+\widetilde{H}^{\mathrm{hyp}},
\end{eqnarray}
where the confinement interaction $\widetilde{H}^{\mathrm{conf}}$ includes a short $\gamma^{\mu}\otimes\gamma_{\mu}$ one-gluon-exchange interaction and an unquenched confining interaction. In the nonrelativistic limit, the effective potential can be written as the standard nonrelativistic expression, i.e.,
\begin{eqnarray}
V_{\mathrm{eff}}(\mathbf{p},\mathbf{r})=H^{\mathrm{conf}}+H^{\mathrm{so}}+H^{\mathrm{hyp}} \label{eq4}
\end{eqnarray}
with
\begin{eqnarray}
\begin{split}
H^{\mathrm{conf}}&=\left[-\frac{3}{4}\left(c+\frac{b(1-e^{-\mu r})}{\mu}\right)+\frac{\alpha(r)}{r}\right]\ (\bm{F_{q}}\cdot\bm{F_{\bar{q}}})\\
&=S(r)+G(r),
\end{split}
\end{eqnarray}
where $H^{\mathrm{conf}}$ contains the spin-independent color screened confinement $S(r)=c+b(1-e^{-\mu r})/\mu$ and Coulomb-type interaction $G(r)=\alpha(r)/r$, and $\bm{F}$ is related to the Gell-Mann matrices in color space, and $\langle\bm{F_{q}}\cdot\bm{F_{\bar{q}}}\rangle=-4/3$ is taken in meson system. The third term $H^{\mathrm{hyp}}$ is the color-hyperfine interaction and reads
\begin{eqnarray}
\begin{split}
H^{\mathrm{hyp}}=&-\frac{\alpha_s(r)}{m_{q}m_{\bar{q}}}\left[\frac{8\pi}{3}\mathbf{S_{q}}\cdot\mathbf{S_{\bar{q}}}\delta^3(\mathbf{r})\right. \\
&\left.+\frac{1}{r^3}\left(\frac{3\mathbf{S_{q}}\cdot\mathbf{r}\mathbf{S_{\bar{q}}}\cdot\mathbf{r}}{r^2}-\mathbf{S_{q}}\cdot\mathbf{S_{\bar{q}}}\right)\right]\ \left(\bm{F_{q}}\cdot\bm{F_{\bar{q}}}\right). \label{eq6}
\end{split}
\end{eqnarray}
The second term $H^{\mathrm{so}}=H^{\mathrm{so(cm)}}+H^{\mathrm{so(tp)}}$ in Eq. (\ref{eq4})
is the spin-orbit interaction with the color magnetic term resulting from one-gluon-exchange
\begin{eqnarray}
H^{\mathrm{so(cm)}}=-\frac{\alpha_s(r)}{r^3}\left(\frac{1}{m_{q}}+\frac{1}{m_{\bar{q}}}\right)\ \left(\frac{\mathbf{S_{q}}}{m_{q}}+\frac{\mathbf{S_{\bar{q}}}}{m_{\bar{q}}}\right)
\cdot\bm{L}\left(\bm{F_{q}}\cdot\bm{F_{\bar{q}}}\right)
\end{eqnarray}
and Thomas precession term
\begin{eqnarray}
H^{\mathrm{so(tp)}}=-\frac{1}{2r}\frac{\partial H^{\mathrm{conf}}}{\partial r}\ \left(\frac{\mathbf{S_{q}}\cdot\bm{L}}{m_{q}^2}+\frac{\mathbf{S_{\bar{q}}}\cdot\bm{L}}{m_{\bar{q}}^2}\right). \label{eq8}
\end{eqnarray}
In the above expressions, $\mathbf{S_{q}}$ and $\mathbf{S_{\bar{q}}}$ denote the spin of light flavor quark and antiquark, respectively, while $\bm{L}$ is the orbital momentum between two quarks inside the meson.

The relativistic effect is introduced by two ways (see Ref. \cite{Godfrey:1985xj} for more details). First, considering the effects of internal motion inside a light hadron and the non-locality interactions between light quark and antiquark, a smearing function could be introduced, {\it i.e.},\begin{eqnarray}
\rho\left(\bm{r}-\bm{r'}\right)=\frac{\sigma_{12}^3}{\pi^{3/2}}e^{-\sigma_{12}^2(\bm{r}-\bm{r'})^2}
\end{eqnarray}
with
\begin{eqnarray}
\begin{split}
\sigma_{12}^2=&\sigma_0^2\left[\frac{1}{2}+\frac{1}{2}\left(\frac{4m_{q}m_{\bar{q}}}{(m_{q}+m_{\bar{q}})^2}\right)^4\right]\\
&+s^2\left(\frac{2m_{q}m_{\bar{q}}}{m_{q}+m_{\bar{q}}}\right)^2,
\end{split}
\end{eqnarray}
where $\sigma_0=1.8$ GeV and $s=3.88$ GeV are constant. Hence, the unquenched confinement potential $S(r)$ and one-gluon exchange interaction $G(r)$ become a smeared potential $\widetilde{S}(r)$ and $\widetilde{G}(r)$ by the general transformation
\begin{eqnarray}\label{sm}
\widetilde{f}(r)=\int d^3r'\rho(\bm{r}-\bm{r'})f(r').
\end{eqnarray}
In addition, a relativized potential should depend on quark momentum, which can be taken into account by introducing momentum-dependent factor. So the Coulomb term $\widetilde{G}(r)$ and the contact, tensor, vector spin-orbital and scalar spin-orbital potential $\widetilde{V}_i(r)$ could be modified as
\begin{eqnarray}
\widetilde{G}(r)\rightarrow\left(1+\frac{p^2}{E_{q}E_{\bar{q}}}\right)^{1/2}\widetilde{G}(r)\left(1+\frac{p^2}{E_{q}E_{\bar{q}}}\right)^{1/2}, \nonumber \\
\frac{\widetilde{V}_i(r)}{m_{q}m_{\bar{q}}}\rightarrow\left(\frac{m_{q}m_{\bar{q}}}{E_{q}E_{\bar{q}}}\right)^{1/2+\epsilon_i}\frac{\widetilde{V}_i(r)}{m_{q}m_{\bar{q}}}
\left(\frac{m_{q}m_{\bar{q}}}{E_{q}E_{\bar{q}}}\right)^{1/2+\epsilon_i},
\end{eqnarray}
where $E_{q}$ ($E_{\bar{q}}$) is the energy of quark (antiquark), and $\epsilon_i$ denotes a momentum correction parameter with a different type of interactions in Eqs. (\ref{eq6})-(\ref{eq8}). One may see that the momentum-dependent factor returns to unity in the nonrelativistic limit.

By diagonalizing the Hamiltonian in Eq. (\ref{eq1}) with a series of the SHO basis, the obtained eigenvalues and eigenvectors can correspond to the meson mass and wave function, respectively.
In the momentum space, the general form of an SHO wave function is
\begin{eqnarray}\label{H}
\begin{split}
\Psi_{nLM_L}(\mathbf{p})=&R_{nL}(p,\beta)Y_{LM_L}(\mathbf{\Omega}_{p})
\end{split}
\end{eqnarray}
with
\begin{eqnarray}\label{H}
\begin{split}
R_{nL}(p,\beta)=&\frac{(-1)^n(-i)^L}{\beta^{3/2}}\;e^{-\frac{p^2}{2\beta^2}}\sqrt{\frac{2n!}{\Gamma(n+L+3/2)}}\left(\frac{p}{\beta}\right)^L\\
&\times L_n^{L+1/2}\left(\frac{p^2}{\beta^2}\right),
\end{split}
\end{eqnarray}
where $R_{nL}(p,\beta)$ is a radial wave function of a harmonic oscillator, and $Y_{LM_L}(\mathbf{\Omega}_p)$ is a spherical harmonic function and $L_n^{L+1/2}(x)$ is the associated Laguerre polynomial.

\section{ Quark pair creation model}\label{appendix-b}

The quark pair creation model \cite{Micu:1968mk,LeYaouanc:1977gm} is usually applied to study the OZI-allowed two-body strong decays of a hadronic state, which are absolutely dominant decay modes for the meson or baryon state above the decay threshold. In the following, we will give a brief introduction of the QPC model.
When a meson decays, a quark-antiquark pair created from the vacuum with the quantum number $J^{PC}=0^{++}$ has a connnection with antiquark and quark inside the initial meson to produce two final mesons. The decay matrix element of this process $A \to BC$ can be written as $\left\langle BC| \mathcal{T} |A \right\rangle=\delta^3(\mathbf{P}_{B}+\mathbf{P}_{C})\times \mathcal{M}^{M_{J_{A}}M_{J_{B}}M_{J_{C}}}(\mathbf{P})$, where the transition operator $\mathcal{T}$ represents a quark-antiquark pair creation from the vacuum. Taking a particle $A$ as an example, the wave function of the mock state $(A,B,C)$ is defined as \cite{Micu:1968mk,LeYaouanc:1977gm}
\begin{eqnarray}\label{H}
&&|A(n_A^{2S_A+1}L_{AJ_AM_{J_A}})(\bm{P}_A)\rangle \equiv \sum\limits_{M_{L_A},M_{S_A}}
\langle L_AM_{L_A}S_AM_{S_A}|J_AM_{J_A}\rangle \nonumber \\
&& \times \sqrt{2E_A} \int d^3\bm{p}_A\psi_{n_AL_AM_{L_A}}(\bm{p}_A)\chi_{S_AM_{S_A}}^{12}\phi_A^{12}\omega_A^{12} \nonumber \\
&& \times \left|q_1\left(\frac{m_1}{m_1+m_2}\bm{P}_A+\bm{p}_A\right)\bar{q}_2\left(\frac{m_2}{m_1+m_2}\bm{P}_A+\bm{p}_A\right)\right\rangle,
\end{eqnarray}
where $m_1$ and $m_2$ are masses of quark $q_1$ and antiquark $\bar{q}_2$, respectively, and $n_A$ is the radial quantum number of a meson $A$, and $\bm{S}_A$ and $\bm{L}_A$  are spin $\bm{S}_q+\bm{S}_{\bar{q}}$ and relative orbital angular momentum between $q_1$ and $\bar{q_2}$, respectively. $\bm{J}_A=\bm{S}_A+\bm{L}_A$ is the total spin, while $\bm{P}_A=\bm{p}_1+\bm{p}_2$ and $E_A$ are center-of-mass (CM) momentum and energy, respectively. $\bm{p}_A=\frac{m_1\bm{p_1}-m_2\bm{p_2}}{m_1+m_2}$ denotes the relative momentum between $q_1$ and $\bar{q}_2$. $\chi_{S_AM_{S_A}}^{12}$, $\phi_{A}^{12}$, $\omega_A^{12}$, and $\psi_{n_AL_AM_{L_A}}(\bm{p}_A)$ are the spin, flavor, color, and spatial wave function of a meson $A$, respectively.
The total decay width of $A \to BC$ in the center-of-mass (CM) frame is given by
\begin{eqnarray}\label{H}
\Gamma_{A \to BC}=\frac{\pi}{4}\frac{|\mathbf{P}|}{m_A^2}\sum\limits_{JL}\left|\mathcal{M}^{JL}(\mathbf{P})\right|^2.
\end{eqnarray}
Here, $\mathbf{P}=\mathbf{P}_{B}=-\mathbf{P}_{C}$. $\bm{L}$ and $\bm{J}$ denote the relative orbital angular and total spin momentum between final states $B$ and $C$, respectively. $\mathcal{M}^{JL}(\mathbf{P})$ is the partial wave amplitude, which can be directly related to the helicity amplitude $\mathcal{M}^{M_{J_A}M_{J_B}M_{J_C}}(\mathbf{P})$ according to the Jacob-Wick formula \cite{Jacob:1959}. In the CM frame the specific form of $\mathcal{M}^{M_{J_A}M_{J_B}M_{J_C}}(\mathbf{P})$ can be written as
\begin{widetext}
\begin{eqnarray}\label{H}
\begin{split}
\mathcal{M}^{M_{J_A}M_{J_B}M_{J_C}}(\mathbf{P})=&\gamma\sqrt{8E_AE_BE_C}\sum\limits_{M_{L_A},M_{S_A},M_{L_B},M_{S_B},M_{L_C},M_{S_C}}\langle L_AM_{L_A}S_AM_{S_A}|J_AM_{J_A}\rangle\;
\langle L_BM_{L_B}S_BM_{S_B}|J_BM_{J_B}\rangle\\
&\times\langle L_CM_{L_C}S_CM_{S_C}|J_CM_{J_C}\rangle\langle 1m1-m|00\rangle\;\langle\chi_{S_BM_{S_B}}^{14}\chi_{S_CM_{S_C}}^{32}|\chi_{S_AM_{S_A}}^{12}\chi_{1-m}^{34}\rangle\\
&\times\left[\langle \phi_B^{14}\phi_C^{32}|\phi_A^{12}\phi_0^{34}\rangle \mathcal{I}(\mathbf{P},m_2,m_1,m_3)+(-1)^{1+S_A+S_B+S_C}\langle \phi_B^{32}\phi_C^{14}|\phi_A^{12}\phi_0^{34}\rangle \mathcal{I}(\mathbf{-P},m_2,m_1,m_3)\right],
\end{split}
\end{eqnarray}
\end{widetext}
where $\gamma$ reflects the creation possibility of a quark pair $q_3\bar{q_4}$ from the vacuum, which is generally considered to be a universal constant for the decay of a specific meson system and can be determined by the relevant experimental data. It needs to be mentioned that the creation strength for the $s\bar{s}$ pair creation is different from that of the $u\bar{u}+d\bar{d}$ pair, where there exists the relation $\gamma_{s}=\gamma_{u}/\sqrt{3}$ \cite{LeYaouanc:1977gm}. Additionally, $m_3$ is the mass of constituent quark $q_3$ and $\phi_0$ is the flavor wave function of $q_3\bar{q_4}$ pair. The expression of the momentum space integral $\mathcal{I}(\mathbf{P},m_2,m_1,m_3)$ reads as
\begin{eqnarray}\label{H}
\begin{split}
\mathcal{I}(\mathbf{P},m_2,m_1,m_3)=&\int d^3\bm{p}\psi_{n_BL_BM_{L_B}}^*(\frac{m_3}{m_1+m_3}\mathbf{P}+\bm{p})\\
&\times \psi_{n_CL_CM_{L_C}}^*(\frac{m_3}{m_2+m_3}\mathbf{P}+\bm{p})\\
&\times \psi_{n_AL_AM_{L_A}}(\mathbf{P}+\bm{p})\mathcal{Y}_1^m(\bm{p}), \\
&
\end{split}
\end{eqnarray}
where $\mathcal{Y}_1^m(\bm{p})$ denotes the solid harmonic polynomial.

\section*{ACKNOWLEDGEMENTS}

The authors would like to thank Dr. Wen-Biao Yan and Ya-Teng Zhang for useful discussions. This work is supported by the China National Funds for Distinguished Young Scientists under Grant No. 11825503, National Key Research and Development Program of China under Contract No. 2020YFA0406400, the 111 Project under Grant No. B20063, and the National Natural Science Foundation of China under Grant No. 12047501.

\end{document}